\documentclass[aps,pre,preprint,superscriptaddress,amsmath,amssymb,nofootinbib]{revtex4-2}

\usepackage{amsmath}
\usepackage{graphicx}
\usepackage{amssymb}
\usepackage{dcolumn}
\usepackage{bm}
\usepackage{color}
\usepackage{float}
\usepackage{upgreek}
\usepackage{gensymb}
\usepackage[final]{hyperref} 
\hypersetup{
	colorlinks=true,       
	linkcolor=blue,        
	citecolor=blue,        
	filecolor=magenta,     
	urlcolor=blue
}
\usepackage{upgreek}
\usepackage[normalem]{ulem}

\makeatletter

\begin{document}
\title{Diffusion of chiral active Brownian particles in oscillatory Poiseuille flow}

\author{Anshika Sharma}
\thanks{Email: Sharmaanshika0901@gmail.com}
\affiliation{Department of Physics and Material Science and Engineering, Jaypee Institute of Information Technology, Noida, Uttar Pradesh 201304, India}

\author{Vikas Sharma}
\thanks{Email: vikassharma12387@gmail.com}
\affiliation{Department of Computer Science and Engineering, Jaypee Institute of Information Technology, Noida, Uttar Pradesh 201304, India}

\author{Narender Khatri}
\thanks{Corresponding author: narenderkhatri8@gmail.com}
\affiliation{Department of Physics and Material Science and Engineering, Jaypee Institute of Information Technology, Noida, Uttar Pradesh 201304, India}

\author{Suneet Kumar Awasthi}
\thanks{Corresponding author: suneet\_electronic@yahoo.com}
\affiliation{Department of Physics and Material Science and Engineering, Jaypee Institute of Information Technology, Noida, Uttar Pradesh 201304, India}

\date{\today}
	
\begin{abstract}

Micro- and nanoswimmers diffusing in a fluid flow confined to microchannels are often described by overdamped active Brownian particle dynamics, in which viscous effects dominate and inertia is negligible. Here, we investigate the diffusive behavior of chiral self-propelled particles in a two-dimensional microchannel with oscillatory Poiseuille flow. Using numerical simulations, we show that the effective diffusion coefficient strongly depends on the oscillation frequency, flow strength, intrinsic angular velocity, and rotational diffusion rate. Interestingly, the effective diffusion coefficient can oscillate with oscillation frequency and flow strength, exhibiting distinct maxima and minima. The enhanced diffusion peaks arise from resonant diffusion associated with the matching between the flow oscillation time scale and the intrinsic characteristic diffusion time scale. Moreover, we identify an additional vorticity matching mechanism, leading to alternating channel wall accumulation of particles during successive flow half-cycles. Additionally, chiral particles strongly amplify the resonant diffusion, producing diffusion enhancements of several orders of magnitude compared to nonchiral particles. The results of this study could have implications for controlling the diffusion of self-propelled particles in microfluidic channels using time-dependent flow. 

\end{abstract}
	
\maketitle

\newpage

\section{Introduction}

During the past two decades, there has been growing research interest in micro- and nanoscale active particles, both biological and synthetic, due to their diverse applications, including targeted drug delivery~\cite{Tang_et_al@SR:2020, Bozuyuk_et_al@ACSN:2018}, cargo transport~\cite{Xu_et_al@Small:2019}, microfluidics~\cite{Woodhouse_Dunkel@NC:2017}, microsurgery~\cite{Vyskocil_et_al@AN:2020}, and numerous others~\cite{Soto_et_al@ACIE:2020,Bunea_et_al@Micromachines:2020}. Active particles, also referred to as self-propelled particles, are entities that continuously consume free energy from their environment and convert it into persistent motion under nonequilibrium conditions. Researchers have extensively studied various propulsion mechanisms and the diverse dynamical properties of these systems~\cite{Gompper_et_al@JPCM:2020,Bechinger_et_al@RMP:2016,Elgeti_et_al@RPP:2015,Cates@RPP:2012,Wang2015,Zoettl2016, Gompper_et_al@JPCM:2025,Illien2017, Khatri_Kapral@Chaos:2024}. A broad class of active systems is inherently chiral due to internal asymmetry or external torque~\cite{Liebchen_Levis@PRL:2017,Kummel_et_al@PRL:2013,Wang_et_al@PRL:2024, Khatri_et_al@PRE:2026}. As a consequence, chiral active particles, in addition to self-propulsion, follow curved trajectories and exhibit transport characteristics that differ substantially from those of their nonchiral counterparts.

In many biological and physical settings, active particles often diffuse in confined fluid flows rather than in a quiescent fluid. Some important examples include sperm cells swimming in fallopian tubes~\cite{Rutllant_et_al@RDA:2005}, pathogens moving through blood vessels~\cite{Conrad_Poling@ARCBE:2018}, transport of blood cells through the vascular system~\cite{Engstler_et_al@Cell:2007}, and micro- and nanorobots programmed for targeted drug delivery~\cite{Wu_et_al@CSR:2020}. Understanding the transport characteristics of active particles in external fluid flows is important from a biological perspective and crucial for designing artificial micro- and nanoswimmers for applications in targeted drug delivery, cargo transport, cell manipulation, particle separation, water remediation, and disease diagnosis~\cite{Bunea_et_al@Micromachines:2020}. The transport characteristics of active particles in Poiseuille flow through microchannels have been intensively studied both experimentally~\cite{Hill_et_al@PRL:2007,Kaya_Koser@PRL:2009,Kaya_Koser@BJ:2012,Kantsler_et_al@eLife:2014,Jing_et_al@SA:2020} and theoretically~\cite{Jing_et_al@SA:2020,Khatri_Burada@PRE:2022,Zottl_Stark@PRL:2012, Ezhilan_Saintillan@JFM:2015,Vennamneni_et_al@JFM:2020,Costanzo_et_al@JPCM:2012,Nash_et_al@PRL:2010,Zottl_Stark@EPJE:2013,Stark@EPJST:2016,Mathijssen_et_al@PRL:2016,Peng_Brady@PRF:2020,Chilukuri_et_al@PF:2015}. These studies show that the coupling between self-propulsion and fluid flow produces two distinct features: upstream swimming and channel wall accumulation.

Although the transport and diffusion of active particles in steady flows have been widely studied, their diffusive behavior in oscillatory flows remains less explored. Oscillatory flows are ubiquitous in natural environments on a variety of scales, from the cardiovascular system to the tidal system, and are generated by a time-periodic pressure gradient or a time-periodic channel wall motion. Recently, Caldag and Bees studied the effect of oscillatory flow on the diffusive behavior of gyrotactic swimmers~\cite{Caldag_Bees@PTRSA:2025}. Using experiments and simulations, they revealed that oscillatory flow can induce interesting dispersion phenomena. Wang et al.~\cite{Wang_et_al@JFM:2025} reported that spherical non-gyrotactic swimmers can exhibit higher or lower diffusivity than passive solutes due to the disruption of cross-streamline migration governed by Jeffery orbits. More recently, Chakraborty et al.~\cite{Chakraborty_et_al@JFM:2026} formulated the diffusion of nonchiral active particles in an oscillatory Poiseuille flow from a continuum perspective using the Smoluchowski equation. Using different numerical methods, they showed that the effective diffusion coefficient can exhibit an oscillatory behavior as a function of the flow oscillation frequency. Notably, the transport characteristics of chiral active particles in an oscillatory flow through microchannels remain unexplored and have not been quantitatively scrutinized.  

In this article, we investigate the diffusive behavior of chiral self-propelled particles in a microfluidic channel under the influence of an oscillatory Poiseuille flow. We use a minimal overdamped Langevin model to describe the dynamics of self-propelled particles, together with sliding-reflecting boundary conditions~\cite{Khatri@FNL:2024,Khatri_Kapral@JCP:2023,Ghosh_et_al@PRL:2013,Reichhardt_Reichhardt@ARCMP:2017} to incorporate their collisions with the channel walls. Our primary objective in this work is to examine how the coupling between self-propulsion and oscillatory fluid flow influences the effective diffusion and spatial distribution of these chiral particles in the microchannel. The article is organized as follows: In Sec.~\ref{sec: Model}, we introduce our model for active particles in a two-dimensional microchannel with oscillatory Poiseuille flow. A detailed investigation of the diffusive behavior of nonchiral and chiral particles is presented in Secs.~\ref{Sec: Nonchiral_particles} and~\ref{Sec: Chiral_particles}, respectively. We provide our main conclusions in Sec.~\ref{Sec: Conclusions}.
 
\section{Model}
\label{sec: Model}

\begin{figure}[hbt]
\centering
\resizebox{0.85\columnwidth}{!}{%
\includegraphics[scale = 1.0]{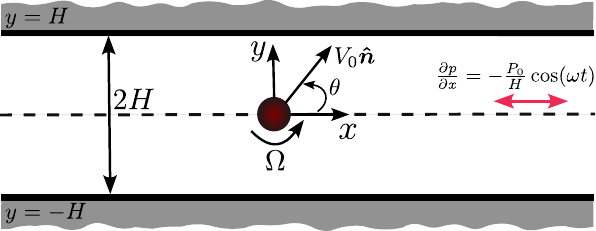}}
\caption{Schematic illustration of a two-dimensional microchannel with oscillatory Poiseuille flow, $\bm{u} = u (y, t) \, \bm{\hat{x}}$, driven by a time-periodic pressure gradient along the $x$ axis, confining the motion of an active (self-propelled) Brownian particle. The width of the channel $2 H$, time-periodic pressure gradient $\partial p/\partial x = - P_{0} \cos (\omega t)/H$, self-propelled velocity $V_{0} \bm{\hat{n}}$, angle $\theta$, and intrinsic angular velocity $\Omega$ are indicated.}  
\label{fig:Model}
\end{figure}

Consider a system of active particles dilute in an incompressible, viscous Newtonian solvent, confined between two parallel walls at $y = \pm H$ with separation $2H$ (see Fig.~\ref{fig:Model}). The active particle density is assumed to be sufficiently low, allowing us to ignore direct particle interactions. We use a minimal overdamped Langevin model for a particle subjected to a one-dimensional flow, $\bm{u} = u (y, t) \, \bm{\hat{x}}$, driven by a prescribed time-periodic pressure gradient along the $x$ axis. The set of coupled Langevin equations for an active particle with position $\bm{r} = (x, y)$, self-propelled velocity $V_{0}$, orientation $\bm{\hat{n}} = (\cos{\theta}, \sin{\theta})$, and intrinsic angular velocity $\Omega$ reads \cite{Hagen_et_al@PRE:2011} 
\begin{equation}\label{Eq:Langevin1}
\begin{aligned}
    \dot{\bm{r}}(t) &= \bm{u} (y, t) + V_{0} \bm{\hat{n}}  + \sqrt{2 D_t}~\bm{\xi} (t), \\ 
    \dot{\theta}(t) &= \Omega_{f} (y, t) + \Omega + \sqrt{2 D_r}~\zeta (t). 
\end{aligned}
\end{equation}
Here $D_t$ and $D_r$ are the translational and rotational diffusion constants, respectively, and $\Omega_{f} (y, t)$ is the local shear rate induced by flow. The random variables $\bm{\xi} (t)$ and $\zeta (t)$ model translational and rotational fluctuations, respectively, arising from the surrounding medium with zero mean and unit variances described by $\langle\boldsymbol{\xi} (t) \otimes \boldsymbol{\xi}^{\mathrm{T}} (t')\rangle = \delta (t - t') \boldsymbol{\mathrm{1}}$ and $\langle \zeta (t) \zeta (t') \rangle = \delta (t - t')$, where $\boldsymbol{\mathrm{1}}$ represents the unit matrix.

The pressure gradient is prescribed by
\begin{equation}\label{Eq:Pressure_Gradient}
\frac{\partial p}{\partial x} = - \frac{P_{0}}{H} \cos (\omega t), 
\end{equation}
where $P_{0}$ is the reference pressure and $\omega$ is the angular frequency of the oscillation.  The Navier–Stokes equations with no-slip boundary conditions at the channel walls yield \cite{Chakraborty_et_al@JFM:2026} $u (y, t) = \mathrm{Re}\left[u'(y)e^{i\omega t}\right]$, where 
\begin{equation}\label{Eq:Velocity_profile1}
u'(y) = \frac{i \, P_{0}}{\rho \, H \,  \omega} \left[-1 + \cosh\left((1+i)\lambda y\right) \mathrm{sech}\left((1+i)\lambda H\right)\right]\!. 
\end{equation}
Here, $\rho$, $\mu$, and $\nu = \mu/\rho$ denote the density, dynamic viscosity, and kinematic viscosity of the fluid, respectively, $i = \sqrt{-1}$ refers to the imaginary unit, and the parameter $1/\lambda = \sqrt{2 \nu/\omega}$ quantifies the viscous length scale over which the fluid momentum diffuses during a period of applied pressure. Defining $U_{f} = P_{0} \, H /(2 \, \mu)$ as the flow strength just for convenience, Eq.~(\ref{Eq:Velocity_profile1}) reads
\begin{equation}\label{Eq:Velocity_profile2}
u'(y) = \frac{i \, U_{f}}{(\lambda \, H)^2} \left[-1 + \cosh\left((1+i)\lambda y\right) \mathrm{sech}\left((1+i)\lambda H\right)\right]\!. 
\end{equation}
In the vanishing frequency limit ($\omega \to 0$), the oscillatory velocity profile reduces to the steady Poiseuille flow as
\begin{equation}\label{Eq:Velocity_profile_SP}
u (y, t) \to U_{f} \left( 1 - \frac{y^2}{H^2} \right)\!. 
\end{equation}
The corresponding vorticity is given by $\Omega_{f} (y, t) = \mathrm{Re}\left[\Omega'(y)e^{i\omega t}\right]$, where
\begin{equation}\label{Eq:Vorticity_profile1}
\Omega'(y) =-\frac{1}{2}\frac{\partial u'}{\partial y}
=\frac{(1-i)U_f}{2\lambda H^2} \sinh\left((1+i)\lambda y\right)
\mathrm{sech}\left((1+i)\lambda H\right)\!.
\end{equation}
As $\omega \to 0$, the vorticity approaches the linear shear profile $\Omega_{f} (y, t) \to U_f \, y/H^2$. 

It is convenient to use a dimensionless description in which all length variables are scaled by channel half-width $H$, $\bm{r^{*}} = \bm{r}/H$, and time by $\tau = H/V_{0}$, $t^{*} = t/\tau$, to work with a constant rescaled self-propelled velocity $V_{0} = 1$. In dimensionless form, Eqs.~(\ref{Eq:Langevin1}) read
\begin{equation}\label{Eq:Langevin2}
\begin{aligned}
    \dot{\bm{r}}(t) &= \bm{u} (y, t) + \bm{\hat{n}}  + \sqrt{2 D_t}~\bm{\xi} (t), \\ 
    \dot{\theta}(t) &= \Omega_{f} (y, t) + \Omega + \sqrt{2 D_r}~\zeta (t), 
\end{aligned}
\end{equation}
where we dispensed with the asterisks for better legibility. Next, flow velocity and vorticity can be written in dimensionless form as $u (y, t) = \mathrm{Re}\left[u'(y)e^{i\chi t}\right]$ and $\Omega_{f} (y, t) = \mathrm{Re}\left[\Omega'(y)e^{i\chi t}\right]$, respectively, where
\begin{equation}\label{Eq:Velocity_profile3}
u'(y) = \frac{i \, u_{f}}{\kappa^2} \left[-1 + \cosh\left((1+i)\kappa y\right) \mathrm{sech}\left((1+i)\kappa\right)\right]\!, 
\end{equation}
\begin{equation}\label{Eq:Vorticity_profile2}
\Omega'(y) =\frac{(1-i)u_f}{2 \kappa} \sinh\left((1+i)\kappa y\right)
\mathrm{sech}\left((1+i)\kappa \right)\!.
\end{equation}
Here $\chi = \omega \tau$ is the dimensionless oscillation frequency, $u_f = U_{f}/V_{0}$ is the dimensionless flow strength, and $\kappa = \lambda H$ compares the viscous length scale $1/\lambda$ with $H$. The other dimensionless parameters are given as follows: $D_{t} \to D_{t}/(H \, V_{0})$, $\Omega \to \Omega \tau$, and $D_{r} \to D_{r} \tau$. As both $\chi$ and $\kappa$ depend on $\omega$, it is reasonable to define the dimensionless parameter $\alpha = \chi/\kappa^2 = 2 \nu \tau/H^2$. In the absence of flow, the dynamics reduces to those of a chiral active particle, whose trajectory approximately combines a circular arc of radius $R_{\Omega} = 1/\lvert \Omega \rvert$ and a persistence length $l_{\rm p} = \tau_{r}$ (as $V_{0} = 1$)~\cite{Bechinger_et_al@RMP:2016, Teeffelen_Lowen@PRE:2008}, where $\tau_{r} = 1/D_{r}$ is the reorientation time. The collisional dynamics of particles at the channel walls is modeled using sliding-reflecting boundary conditions~\cite{Reichhardt_Reichhardt@ARCMP:2017, Khatri_Burada@PRE:2022, Khatri_Kapral@JCP:2023, Ghosh_et_al@PRL:2013, Khatri@FNL:2024}. More specifically, when a particle encounters the channel wall, its translational velocity $\dot{\bm{r}}$ is elastically reflected while its orientation $\bm{\hat{n}}$ is preserved during impact; therefore, it essentially slides along the channel wall until orientational fluctuations redirect it toward the interior of the channel. Unless stated otherwise, we set $\alpha = 100$ and $D_{t}/D_{r} = 0.1$ throughout this work. 

Unfortunately, the Fokker-Planck equations corresponding to the set of coupled Langevin equations~(\ref{Eq:Langevin2}) are analytically intractable because Eqs.~(\ref{Eq:Langevin2}) become nonlinear in the presence of flow. For this reason, we use Brownian dynamics simulations in the channel performed by the integration of coupled Langevin equations~(\ref{Eq:Langevin2}) using the standard stochastic Euler algorithm \cite{Toschi_Sega@Book:2019, Volpe_et_al@AJP:2014} over $2 \times 10^{4}$ trajectories with a time step smaller than $10^{-4}$. At $t = 0$, the particles are uniformly distributed within a single cell of the channel that spans $x \in [0, 1]$ with random orientations. Since the flow velocity $u (y, t)$ becomes zero upon time-averaging over one period of the flow oscillation, the average velocity of particles vanishes in the asymptotic long-time regime. Therefore, transport characteristics of particles are only characterized by the effective diffusion coefficient $D_{\rm eff}$ along the longitudinal axis ($x$) as the particles are confined in the transverse ($y$) direction. In the long-time limit, $D_{\rm eff}$ along the $x$ direction is numerically calculated as 
\begin{equation}\label{eq:TP_calculation}
D_{\mathrm{eff}}  = \lim_{t\to\infty} \frac{\langle x^2(t) \rangle - \langle x(t) \rangle ^2 }{2 \,t}. 
\end{equation}
In the following sections, we present results for the effective diffusion coefficient and other relevant properties of nonchiral and chiral particles. 
    
\section{Effective Diffusion of Nonchiral Particles}\label{Sec: Nonchiral_particles}

\begin{figure}[htb!]
	\centering	
	\includegraphics[scale = 0.85]{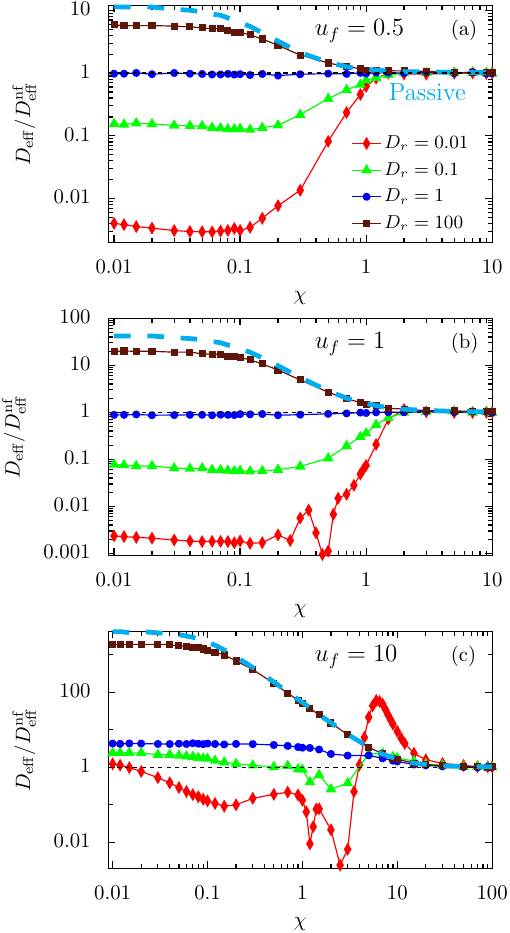}
	\caption{Scaled effective diffusion coefficient $D_{\rm eff}/D_{\rm eff}^{\rm nf}$ as a function of  $\chi$ for different values of $D_{r}$. (a) For $u_{f} = 0.5$, (b) for $u_{f} = 1$, and (c) for $u_{f} = 10$. Here and below, solid lines guide the eye, and the statistical error for $D_{\rm eff}/D_{\rm eff}^{\rm nf}$ is smaller than the symbol size. The other parameters are $\alpha = 100$, $D_{t}/D_{r} = 0.1$, and $\Omega = 0$. The dashed cyan line denotes results for the passive Brownian particles. For the comparison of $D_{\rm eff}/D_{\rm eff}^{\rm nf}$ between passive and self-propelled particles, we use $D_{t} = 0.01$ and $D_r = 100$.}
	\label{fig:Transport_properies_Graph1}
\end{figure}

Although Chakraborty et al.~\cite{Chakraborty_et_al@JFM:2026} recently investigated the diffusion of nonchiral self-propelled particles in a channel with oscillatory Poiseuille flow, we briefly examine their diffusive behavior here for completeness. Figure~\ref{fig:Transport_properies_Graph1} shows the dependence of the scaled effective diffusion coefficient $D_{\rm eff}/D_{\rm eff}^{\rm nf}$ on the oscillation frequency $\chi$ for different values of the rotational diffusion rate $D_{r}$ and at various flow strengths $u_{f}$. In the very high-frequency limit, i.e., $\chi \to \infty$, the fluid advection becomes ineffective due to rapid reversal of the oscillatory flow; thereby causing the effective diffusion coefficient to converge to its no-flow value $D_{\rm eff}^{\rm nf} = D_{t} + \tau_{r}/2$~\cite{Howse_et_al@PRL:2007} ($D_{\rm eff}/D_{\rm eff}^{\rm nf} \to 1$) irrespective of $u_{f}$ or $D_{r}$. For comparison, $D_{\rm eff}/D_{\rm eff}^{\rm nf}$ of passive Brownian particles decreases monotonically with increasing $\chi$ until it reaches the no-flow limit (dashed cyan line). In contrast, for self-propelled particles, $D_{\rm eff}/D_{\rm eff}^{\rm nf}$ displays a markedly richer behavior that strongly depends on $D_{r}$ and $u_{f}$. Particularly when $D_{r} \to \infty$, where re-orientation proceeds rapidly, the self-propelled dynamics effectively reduces to that of passive Brownian particles, and consequently $D_{\rm eff}/D_{\rm eff}^{\rm nf}$ approaches the passive particle result. For various values of $u_{f}$, the combined action of fluid flow and self-propelled motion significantly suppresses $D_{\rm eff}/D_{\rm eff}^{\rm nf}$ at low values of $\chi$ for small $D_{r}$. As $\chi$ increases, $D_{\rm eff}/D_{\rm eff}^{\rm nf}$ monotonically increases for weak flow strengths ($u_{f} < V_{0}$), eventually approaching the no-flow limit (see Fig.~\ref{fig:Transport_properies_Graph1}(a)). Interestingly, for moderate and larger values of $u_{f}$, $D_{\rm eff}/D_{\rm eff}^{\rm nf}$ exhibits an oscillatory dependence on $\chi$ only for small values of $D_{r}$ (see Figs.~\ref{fig:Transport_properies_Graph1}(b)-(c)), indicating the onset of resonant diffusion arising from the matching between the flow oscillation time scale, $\tau_{\chi} \sim 1/\chi$, and an intrinsic time scale of self-propelled particles.
Note that the dynamics of self-propelled particles is governed by several characteristic time scales, including swimming time $H/V_{0}$, reorientation time $1/D_{r}$, characteristic diffusion time $H^2/D_{t}$, flow advection time $H/u_{f}$, and, for chiral particles, intrinsic rotation time  $1/\lvert \Omega \rvert$.
Since these time scales are nonlinearly coupled, the intrinsic time scale associated with resonant diffusion cannot be uniquely identified; however, the observed dependence of the resonance on $u_{f}$ can be rationalized. Compared to $u_{f} = 1$, increasing the flow strength to  $u_{f} = 10$ shifts the onset of oscillatory behavior and the extrema to higher values of $\chi$. This is because the flow advection time $\tau_{f} = H/u_{f}$ decreases with increasing $u_{f}$ and therefore a smaller value of $\tau_{\chi}$ (or a higher value of $\chi$) is required to match $\tau_{f}$. Formally, we write the intrinsic characteristic time as $\tau_{i} = \tau_{i}(V_{0}, D_{r}, D_{t}, u_{f}, \lvert \Omega \rvert)$, which must become comparable with $\tau_{\chi}$ to attain resonance. A similar dependence of $D_{\rm eff}/D_{\rm eff}^{\rm nf}$ on $\chi$ is recently reported in Ref.~\cite{Chakraborty_et_al@JFM:2026}. 

\begin{figure}[htb!]
	\centering	
	\includegraphics[scale = 0.85]{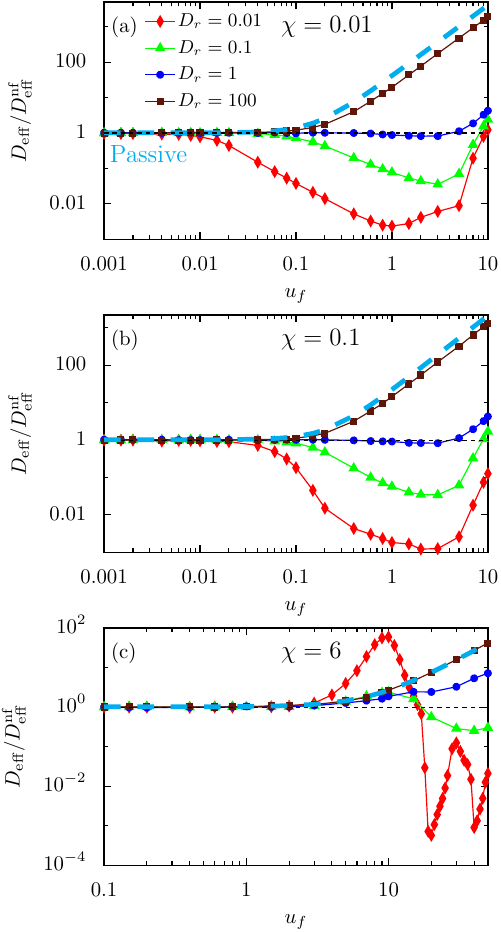}
	\caption{Scaled effective diffusion coefficient $D_{\rm eff}/D_{\rm eff}^{\rm nf}$ as a function of $u_{f}$ for different values of $D_{r}$. (a) For $\chi = 0.01$, (b) for $\chi = 0.1$, and (c) for $\chi = 6$. The other parameters are $\alpha = 100$, $D_{t}/D_{r} = 0.1$, and $\Omega = 0$. The dashed cyan line denotes results for the passive Brownian particles. For the comparison of $D_{\rm eff}/D_{\rm eff}^{\rm nf}$ between passive and self-propelled particles, we use $D_{t} = 0.01$ and $D_r = 100$.}
	\label{fig:Transport_properies_Graph2}
\end{figure}

Figure~\ref{fig:Transport_properies_Graph2} depicts the dependence of the scaled effective diffusion coefficient $D_{\rm eff}/D_{\rm eff}^{\rm nf}$ on flow strength $u_{f}$ for different values of the rotational diffusion rate $D_{r}$ and at various oscillation frequencies $\chi$. In the limit $u_{f} \ll V_{0}$, the fluid velocity can be neglected, and $D_{\rm eff}/D_{\rm eff}^{\rm nf}$ approaches its no-flow value ($D_{\rm eff}/D_{\rm eff}^{\rm nf} \to 1$). For passive Brownian particles, $D_{\rm eff}/D_{\rm eff}^{\rm nf}$ increases monotonically with $u_{f}$. In contrast, self-propelled particles show a markedly different dependence on $u_{f}$ that depends strongly on $\chi$ and $D_{r}$. At low and moderate values of $\chi$, $D_{\rm eff}/D_{\rm eff}^{\rm nf}$ first decreases, reaches a minimum at intermediate $u_{f}$, and subsequently increases with $u_{f}$ for small and moderate values of $D_{r}$ (see Figs.~\ref{fig:Transport_properies_Graph2}(a)-(b)), reflecting that the combined action of fluid flow and self-propelled motion significantly suppresses $D_{\rm eff}/D_{\rm eff}^{\rm nf}$. This result is consistent with the findings in Refs.~\cite{Khatri_Burada@PRE:2022, Peng_Brady@PRF:2020}.
More interestingly, at higher values of $\chi$, $D_{\rm eff}/D_{\rm eff}^{\rm nf}$ displays an oscillatory behavior as a function of $u_{f}$ only for small values of $D_r$ (see Fig.~\ref{fig:Transport_properies_Graph2}(c)), with successive maxima and minima corresponding to alternating enhancement and suppression of diffusion, respectively.  This oscillatory dependence further supports resonant diffusion and indicates that varying $u_{f}$ can tune the resonance. When $u_{f} \gg V_{0}$, the fluid velocity dominates over the self-propelled motion; thus, in this regime, $D_{\rm eff}/D_{\rm eff}^{\rm nf}$ approaches the passive particle result. It should be noted that for smaller values of $D_{r}$ or higher values of $\chi$, a large value of $u_{f}$ is required for the convergence of $D_{\rm eff}/D_{\rm eff}^{\rm nf}$ to the passive particle result. As $D_{r} \to \infty$, $D_{\rm eff}/D_{\rm eff}^{\rm nf}$ approaches the passive particle result, indicating that self-propelled dynamics effectively reduces to passive Brownian motion.

\begin{figure}[htb!]
\centering
\resizebox{1.0\columnwidth}{!}{%
\includegraphics[scale = 1.0]{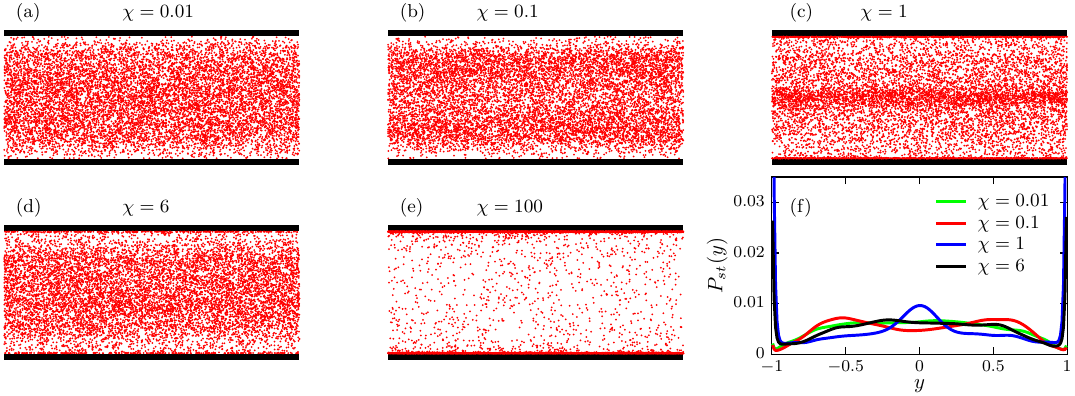}}
\caption{(a)-(e) Steady-state distribution of nonchiral particles for different values of $\chi$. (f) Corresponding normalized probability distributions $P_{st} (y)$ along the $y$ direction. The other parameters are $\alpha = 100$, $D_{r} = 0.01$, $D_{t}/D_{r} = 0.1$, $u_{f} = 10$, and $\Omega = 0$.}
\label{fig:Distribution_Graph1}
\end{figure}

To gain further insight into the oscillatory variation of the scaled effective diffusion coefficient $D_{\rm eff}/D_{\rm eff}^{\rm nf}$ with the oscillation frequency $\chi$ observed in Fig.~\ref{fig:Transport_properies_Graph1}, we consider the steady-state spatial distribution of particles for the same parameters used for the red curve in Fig.~\ref{fig:Transport_properies_Graph1}(c). Figure~\ref{fig:Distribution_Graph1} depicts the steady-state distribution of particles and the corresponding normalized probability distribution $P_{st} (y)$ along the $y$ direction for different values of $\chi$. At low frequencies, i.e., $\chi \to 0$, the combined action of fluid flow and self-propelled motion leads to an approximately uniform distribution of particles across the channel, with a monotonic decay near the channel walls. As expected for nonchiral particles, the particle distribution is symmetric about the principal axis of the channel. As $\chi$ increases, the particle distribution shows a pronounced dependence on $\chi$. At $\chi = 0.1$, particles are preferentially distributed away from the principal axis of the channel, whereas at $\chi = 1$, they accumulate predominantly at the channel walls, accompanied by a central maximum. These distributions suggest that particles at $\chi = 0.01$ cross the local fluid streamlines more rapidly than those at $\chi = 0.1$ and $\chi = 1$, consistent with the relatively higher
$D_{\rm eff}/D_{\rm eff}^{\rm nf}$ at $\chi = 0.01$ (see the red curve in Fig.~\ref{fig:Transport_properies_Graph1}(c)). It should be noted that changes in the particle distribution are accompanied by oscillations of $D_{\rm eff}/D_{\rm eff}^{\rm nf}$ in the intermediate-frequency regime, where successive maxima correspond to the resonance condition and the intervening minima indicate suppressed diffusion.
Interestingly, at $\chi = 6$, the particles predominantly accumulate at the channel walls, while the remaining particles are approximately uniformly distributed throughout the channel except near the channel walls (see the $P_{st} (y)$ plot for $\chi = 6$).
This distinct spatial distribution occurs at the same oscillation frequency at which $D_{\rm eff}/D_{\rm eff}^{\rm nf}$ reaches its maximum in Fig.~\ref{fig:Transport_properies_Graph1}(c), providing further evidence of resonant diffusion associated with the matching between $\tau_{\chi}$ and $\tau_{i}$. In the limit $\chi \to \infty$, most of the particles accumulate at the channel walls, whereas a smaller fraction remains in the channel bulk, reflecting the diminishing contribution of fluid advection during rapid flow oscillations.

\begin{figure}[htb!]
\centering
\resizebox{1.0\columnwidth}{!}{%
\includegraphics[scale = 1.0]{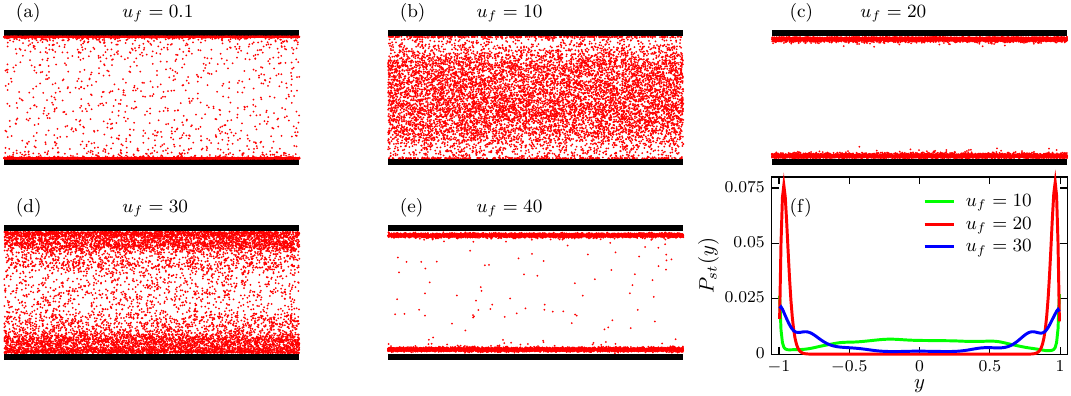}}
\caption{(a)-(e) Steady-state distribution of nonchiral particles for different values of $u_{f}$. (f) Corresponding normalized probability distributions $P_{st} (y)$ along the $y$ direction. The other parameters are $\alpha = 100$, $D_{r} = 0.01$, $D_{t}/D_{r} = 0.1$, $\chi = 6$, and $\Omega = 0$.}
\label{fig:Distribution_Graph2}
\end{figure}
   
To better understand the oscillatory behavior of the scaled effective diffusion coefficient $D_{\rm eff}/D_{\rm eff}^{\rm nf}$ as a function of flow strength $u_{f}$ observed in Fig.~\ref{fig:Transport_properies_Graph2}(c), we analyze the steady-state spatial distribution of particles for the same parameters corresponding to the red curve in that figure. The steady-state distribution of particles and the corresponding normalized probability distribution $P_{st} (y)$ along the $y$ direction are shown in Fig.~\ref{fig:Distribution_Graph2} for different values of $u_{f}$. For small values of $u_{f}$, the distribution of particles is similar to the distribution depicted in Fig.~\ref{fig:Distribution_Graph1}(e), indicating that, in this regime of $u_f$, the very high-frequency limit is achieved for $\chi = 6$ and the fluid advection becomes ineffective. As $u_{f}$ increases, the distribution of particles changes markedly. For $u_{f} = 10$, the particle distribution, as discussed in the description of the previous figure, is a signature of the onset of resonant diffusion, resulting in an enhanced peak in $D_{\rm eff}/D_{\rm eff}^{\rm nf}$. When the flow strength increases to $u_{f} = 20$, the particles become strongly localized near the channel walls (see the $P_{st} (y)$ plot for $u_{f} = 20$), substantially suppressing $D_{\rm eff}/D_{\rm eff}^{\rm nf}$ by restricting their crossing of the local fluid streamlines. On further increasing the flow strength to $u_{f} = 30$, a substantial fraction of the particles redistributes toward the channel interior; however, for $u_{f} = 40$, the particle distribution becomes localized near the channel walls again (see Figs.~\ref{fig:Distribution_Graph2}(d) and \ref{fig:Distribution_Graph2}(e)). These successive changes in the particle distribution are accompanied by oscillations in $D_{\rm eff}/D_{\rm eff}^{\rm nf}$ as a function of $u_{f}$, with successive maxima arising from the resonance condition and the intervening minima resulting from suppressed diffusion.

\section{Effective Diffusion of Chiral Particles}\label{Sec: Chiral_particles}

\begin{figure}[htb!]
	\centering	
	\includegraphics[scale = 0.85]{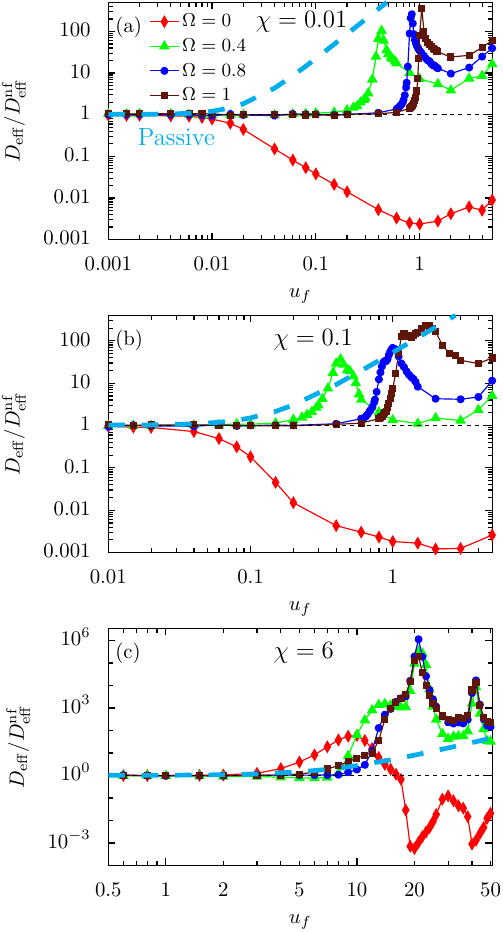}
	\caption{Scaled effective diffusion coefficient $D_{\rm eff}/D_{\rm eff}^{\rm nf}$ as a function of $u_{f}$ for different values of $\Omega$. (a) For $\chi = 0.01$, (b) for $\chi = 0.1$, and (c) for $\chi = 6$. The other parameters are $\alpha = 100$, $D_{r} = 0.01$, and $D_{t}/D_{r} = 0.1$. The dashed cyan line denotes results for the passive Brownian particles.}
	\label{fig:Transport_properies_Graph3}
\end{figure} 

In this section, we examine the diffusive behavior of chiral self-propelled particles. Figure~\ref{fig:Transport_properies_Graph3} depicts the scaled effective diffusion coefficient $D_{\rm eff}/D_{\rm eff}^{\rm nf}$ as a function of the flow strength $u_{f}$ for different values of the angular velocity $\Omega$ and at various oscillation frequencies $\chi$. At low and moderate values of $\chi$, the qualitative behavior of $D_{\rm eff}/D_{\rm eff}^{\rm nf}$ is similar to that reported by Khatri and Burada~\cite{Khatri_Burada@PRE:2022} for different values of $\Omega$. For $u_{f} \ll V_{0}$, as mentioned earlier, the fluid velocity can be neglected; thus, $D_{\rm eff}$ converges to its no-flow value $D_{\rm eff}^{\rm nf} = D_{t} + (\tau_{r}/2)/[1 + (\Omega \tau_{r})^2]$~\cite{Teeffelen_Lowen@PRE:2008,Ebbens_et_al@PRE:2010} ($D_{\rm eff}/D_{\rm eff}^{\rm nf} \to 1$) irrespective of $\chi$ or $\Omega$. In the opposite regime, $u_{f} \gg V_{0}$, the fluid velocity dominates the self-propelled motion, and $D_{\rm eff}/D_{\rm eff}^{\rm nf}$ approaches the passive result (dashed line). Note that this asymptotic regime of very large $u_{f}$ has not yet been reached in Fig.~\ref{fig:Transport_properies_Graph3}. Interestingly, in the low-frequency limit, i.e., $\chi \to 0$, the effect of the local fluid vorticity $\Omega_{f}$ near the lower channel wall is canceled out when $\Omega \approx u_{f}$ during the positive half-cycle of the flow velocity, leading to the accumulation of particles near the lower wall. During the negative half-cycle of the flow velocity, the same mechanism operates near the upper channel wall. Consequently, $D_{\rm eff}/D_{\rm eff}^{\rm nf}$ exhibits an enhanced diffusion peak (see Fig.~\ref{fig:Transport_properies_Graph3}(a)). For $\Omega < 0$, the same mechanism occurs at the opposite channel wall, and we have verified that $D_{\rm eff}/D_{\rm eff}^{\rm nf}$ is independent of the sign of $\Omega$. As $\chi$ increases to moderate values, the enhanced diffusion peak associated with the match between $\Omega$ and $\Omega_{f}$ near a channel wall becomes broader, less pronounced and slightly shifts toward larger values of $u_{f}$. At higher values of $\chi$, this mechanism becomes ineffective, and the corresponding enhanced diffusion peak disappears. Rather, $D_{\rm eff}/D_{\rm eff}^{\rm nf}$ develops an oscillatory behavior as a function of $u_{f}$, with successive maxima corresponding to the resonant diffusion arising from the temporal matching between $\tau_{\chi}$ and $\tau_{i}$ and the intervening minima representing suppressed diffusion. Compared with nonchiral particles, these oscillations are more pronounced for chiral particles, and the enhancement in $D_{\rm eff}/D_{\rm eff}^{\rm nf}$ exceeds the no-flow value by several orders of magnitude (see Fig.~\ref{fig:Transport_properies_Graph3}(c)). Additionally, the positions of the maxima and minima in $D_{\rm eff}/D_{\rm eff}^{\rm nf}$ for nonchiral and chiral particles differ. 

\begin{figure}[htb!]
	\centering	
	\includegraphics[scale = 1.0]{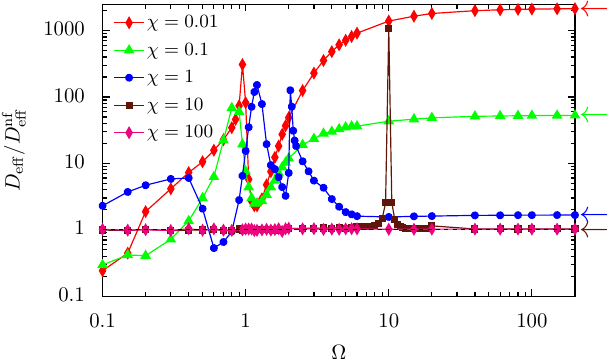}
	\caption{Scaled effective diffusion coefficient $D_{\rm eff}/D_{\rm eff}^{\rm nf}$ as a function of $\Omega$ for different values of $\chi$ at fixed flow strength $u_{f} = 1$. The other parameters are $\alpha = 100$, $D_{r} = 0.01$, and $D_{t}/D_{r} = 0.1$. The left arrows denote results for the passive Brownian particles.}
	\label{fig:Transport_properies_Graph4}
\end{figure}

Figure~\ref{fig:Transport_properies_Graph4} shows the scaled effective diffusion coefficient $D_{\rm eff}/D_{\rm eff}^{\rm nf}$ as a function of the angular velocity $\Omega$ for different values of the oscillation frequency $\chi$ at fixed flow strength $u_{f} = 1$. At low values of $\chi$, $D_{\rm eff}/D_{\rm eff}^{\rm nf}$ exhibits an enhanced diffusion peak around $\Omega \simeq 1$ arising from the matching between $\Omega$ and the local flow-induced vorticity $\Omega_f$ near a channel wall. Interestingly, as $\chi$ increases to moderate values, an additional pronounced peak emerges at larger $\Omega$ due to resonant diffusion arising from the matching between $\tau_{\chi}$ and $\tau_{i}$. Upon further increasing $\chi$, the enhanced diffusion peak associated with the matching between $\Omega$ and $\Omega_{f}$ near the channel wall disappears, while the resonant diffusion peak shifts toward larger $\Omega$. Note that as $\Omega \to \infty$, the motion of self-propelled particles effectively reduces to that of passive Brownian particles; thus, $D_{\rm eff}/D_{\rm eff}^{\rm nf}$ approaches the passive particle result. In the very high-frequency limit, i.e., $\chi \to \infty$, as expected, $D_{\rm eff}$ converges to its no-flow value ($D_{\rm eff}/D_{\rm eff}^{\rm nf} \to 1$). 

\begin{figure}[htb!]
\centering
\resizebox{1.0\columnwidth}{!}{%
\includegraphics[scale = 1.0]{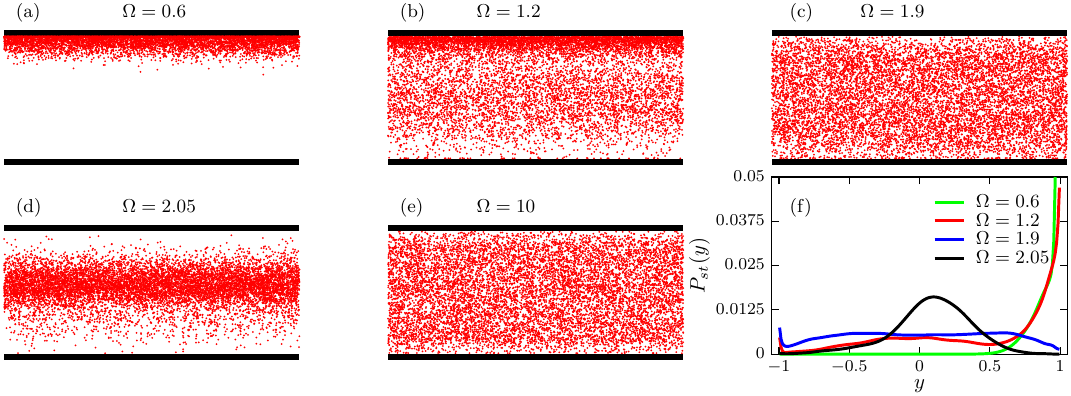}}
\caption{(a)-(e) Steady-state distribution of chiral particles for different values of $\Omega$. (f) Corresponding normalized probability distributions $P_{st} (y)$ along the $y$ direction. The other parameters are $\alpha = 100$, $D_{r} = 0.01$, $D_{t}/D_{r} = 0.1$, $\chi = 1$, and $u_{f} = 1$.}
\label{fig:Distribution_Graph3}
\end{figure}

To further elucidate the observed behavior of the scaled effective diffusion coefficient $D_{\rm eff}/D_{\rm eff}^{\rm nf}$ as a function of the angular velocity $\Omega$, we consider the steady-state distribution of chiral particles for the same parameters used for the blue curve in Fig.~\ref{fig:Transport_properies_Graph4}. Figure~\ref{fig:Distribution_Graph3} shows the steady-state distribution of particles and the corresponding normalized probability distribution $P_{st}(y)$ along the $y$ direction for different values of $\Omega$. When $\Omega \approx 1$, most of the particles accumulate near the upper channel wall due to the match between $\Omega$ and local flow-induced vorticity $\Omega_f$ near the upper channel wall. The same will happen near the lower channel wall during the subsequent half-cycle of the flow velocity. This mechanism produces the first enhanced diffusion peak at $\Omega = 1.2$. As $\Omega$ increases to $\Omega = 1.9$, the channel wall accumulation disappears, and the particle distribution and the corresponding $P_{st} (y)$ become nearly uniform across the channel. Interestingly, at $\Omega = 2.05$, corresponding to the second enhanced diffusion peak in Fig.~\ref{fig:Transport_properies_Graph4} (see the blue curve), most of the particles accumulate slightly above the principal axis of the channel. Accordingly, $P_{st} (y)$ attains its maximum slightly above the principal axis of the channel and decreases monotonically away from this location. Note that during the subsequent half-cycle of the flow velocity, particles accumulate symmetrically slightly below the principal axis of the channel. This distinct redistribution of particles provides evidence of resonant diffusion arising from the matching between $\tau_{\chi}$ and $\tau_{i}$. As one would expect, in the limit $\Omega \to \infty$, the particle distribution and the corresponding $P_{st} (y)$ become uniform throughout the channel.

\section{Concluding Remarks}\label{Sec: Conclusions}

In this work, we have studied the diffusive behavior of both nonchiral and chiral self-propelled particles confined in a two-dimensional microchannel with oscillatory Poiseuille flow. Using Brownian dynamics simulations, we have shown that the scaled effective diffusion coefficient exhibits peculiar features depending on the oscillation frequency $\chi$, flow strength $u_{f}$, intrinsic angular velocity $\Omega$, and rotational diffusion rate $D_{r}$. For nonchiral particles, we have observed that the scaled effective diffusion coefficient can oscillate as a function of $\chi$ and $u_{f}$. The enhanced diffusion peaks are associated with a resonance between the flow oscillation time scale $\tau_{\chi}$ and the intrinsic characteristic time scale $\tau_{i}$. Interestingly, for chiral particles, we have demonstrated an additional mechanism for enhanced diffusion at low and moderate values of $\chi$, originating from the matching between the intrinsic angular velocity and the local flow-induced vorticity near a channel wall. In particular, this vorticity matching mechanism leads to preferential accumulation of particles near the lower channel wall during the positive half-cycle of the flow velocity and near the upper channel wall during the subsequent negative half-cycle, resulting in an enhanced diffusion peak. As $\chi$ increases further, this mechanism gradually weakens, and the corresponding enhanced diffusion peak disappears. Moreover, it is found that chiral particles show strongly enhanced resonant diffusion peaks compared to nonchiral particles, with diffusion enhancements of several orders of magnitude.
Additionally, the positions of maxima and minima in the scaled effective diffusion coefficient differ between nonchiral and chiral particles. The ability to effectively control the diffusion of self-propelled particles in oscillatory flow by varying the oscillation frequency and flow strength has significant implications for a wide range of applications, including the design of high-efficiency bioreactors~\cite{Ober_et_al@PNAS:2015, Gogate_et_al@BEJ:2000}, targeted drug delivery~\cite{Bozuyuk_et_al@ACSN:2018,Gao_et_al@Small:2012}, microfluidic and lab-on-a-chip technologies~\cite{Mudugamuwa_et_al@Small:2024,Dittrich_Manz@NRDD:2006}, control of biofilm formation~\cite{Conrad_Poling@ARCBE:2018,Karimi_et_al@LC:2015}, and many more. 

\section{Acknowledgments}

This work was supported by the Jaypee Institute of Information Technology, Noida. The authors gratefully acknowledge the Ramanujan Universe HPC facility at Jaypee Institute of Information Technology, Noida, for providing the computational resources.


\bibliography{APS} 



\end{document}